\documentclass[12pt,a4paper]{article}
\usepackage[left=2.3cm,top=2.3cm,right=2.3cm,bottom=2.3cm]{geometry}
\usepackage[utf8x]{inputenc}
\usepackage{ucs}
\usepackage{authblk}
\usepackage{amsmath,amsfonts,bm,amssymb,latexsym,amsthm}
\usepackage{graphicx, graphics}
\usepackage{multirow}
\usepackage[numbers,sort&compress]{natbib}
\usepackage{microtype}
\usepackage[none]{hyphenat}
\usepackage{setspace} 
\usepackage{array}
\usepackage{amsfonts}
\usepackage[bottom]{footmisc}
\usepackage{color}
\usepackage{xurl}
\usepackage{color,soul}

\usepackage{float} 
\usepackage{rotating}
\usepackage{morefloats}
\usepackage{booktabs}
\usepackage{multirow}
\usepackage[para,online,flushleft]{threeparttable}
\usepackage[justification=justified,font=small]{caption}
\usepackage{hyperref}
\hypersetup{colorlinks, breaklinks, linkcolor=[rgb]{0,0.3,1}, citecolor=[rgb]{0,0.3,1}, urlcolor=[rgb]{0,0.3,1} }
\usepackage{longtable}
\usepackage{multirow}
\usepackage{subcaption}
\usepackage{lscape}

\title{Changes in mobility and socioeconomic conditions in Bogot\'a city during the COVID-19 outbreak}
\author[1,*]{Marco Due\~nas} 
\author[2]{Mercedes Campi}
\author[3]{Luis Olmos}

\affil[1]{\small Department of Economics, International Trade and Social Policy -- Universidad de Bogot\'a Jorge Tadeo Lozano, Bogot\'a Colombia}
\affil[2]{\small CONICET - University of Buenos Aires, Faculty of Economics, IIEP, Buenos Aires, Argentina}
\affil[3]{\small Department of City and Regional Planning, University of California, Berkeley, CA 94720, United States}
\affil[*]{\small Corresponding author: maduenase@gmail.com}

\begin{document}
\maketitle

\begin{abstract}
We analyze mobility changes following the implementation of containment measures aimed at mitigating the spread of COVID-19 in Bogot\'a, Colombia. We characterize the mobility network before and during the pandemic and analyze its evolution and changes between January and July 2020. We then link the observed mobility changes to socioeconomic conditions, estimating a gravity model to assess the effect of socioeconomic conditions on mobility flows. We observe an overall reduction in mobility trends, but the overall connectivity between different areas of the city remains after the lockdown, reflecting the mobility network's resilience. We find that the responses to lockdown policies depend on socioeconomic conditions. Before the pandemic, the population with better socioeconomic conditions shows higher mobility flows. Since the lockdown, mobility presents a general decrease, but the population with worse socioeconomic conditions shows lower decreases in mobility flows. We conclude deriving policy implications.
\end{abstract}

\medskip 
\noindent \textbf{Keywords:} Mobility networks; Poverty; Informality; Socioeconomic strata; COVID-19

\clearpage
\newpage
\section{Introduction}

Since the outbreak of the 2019 novel coronavirus (COVID-19) pandemic, governments have been implementing containment measures aimed at mitigating the spread of the virus. Several policies restricted human mobility, intending to increase social distance, which has been effective in slowing the transmission. The ability to adapt to the pandemic and respond to containment measures can be bound by socioeconomic conditions, which are heterogeneous in large urban areas of low- and middle-income countries.

There is broad agreement that aggregated mobility data could help fight COVID-19 by providing metrics for data-supported decisions and informing public health actions \citep{buckee2020aggregated, oliver2020mobile}. Mostly using large scale mobile phone data or social media data, several studies show that mobility significantly changed during lockdowns in high-income countries \citep{badr2020association, gatto2020spread, linka2020outbreak, gao2020mapping}. Other studies provide evidence for middle-income countries like China and Brazil \citep{chinazzi2020effect, kraemer2020effect, coelho2020}.
%, such as the United States, Germany, Italy, Japan, France, Spain, Sweden, United Kingdom, and in Europe %Mobility WP: \citep{klein2020, schlosser2020covid, yabe2020non, pepe2020, bonato2020mobile, wellenius2020impacts, dahlberg2020effects, santana2020, cintia2020relationship}
%China lai2020,

In addition to the impact on health, the COVID-19 generates socioeconomic effects of still unknown magnitude \citep{nicola2020socio, world_bank2020}. The crisis has severely affected labor, increasing unemployment rates, and decreasing hours of work and labor force participation around the world \citep{beland2020short, del2020supply}. These adverse effects are expected to be unevenly distributed between and within countries. Inequality could spread COVID-19, and, at the same time, the pandemic could exacerbate inequality \citep{ahmed2020inequality}. Likewise, responses to policies and measures to contain COVID-19 could be different due to existing inequality \citep{blundell2020covid}.

In Colombia, ``lockdown" policies included border closures and travel bans, public activity restrictions, and school and business closures. The effect of these policies on urban mobility has not been addressed yet. We present evidence for Bogot\'a, the major urban area, and capital of Colombia, with 7.41 million inhabitants, geographically and administratively divided into 20 administrative districts --named Localities. Although there have been improvements in recent years, Bogot\'a is still a city with heavy traffic congestion and a large and chaotic transportation system \citep{olmos2004cellular}. The public transport network has over 2.6 million users per day on average and suffers heavy congestion, particularly in rush hours and some locations. The city is also characterized by high levels of income inequality, poverty, and labor informality. 
%COMO SE DICE HORAS PICO?

%https://colombiareports.com/colombia-poverty-inequality-statistics/
%Me gustaria caracterizar el trafico de bogota en una frase, quizas esto es viejo y se puede poner algo mas actualizado

We analyze mobility flows collected from smart card validations at the integrated public transport system. We contribute with a characterization of the mobility network before and after the implementation of policies restricting human mobility in a large urban area of a middle-income country. Then, we link changes in mobility flows to socioeconomic conditions to understand if there have been different responses to mobility restrictions.

We observe a general decrease in mobility flows following social distancing interventions on population mobility. However, responses to lockdown policies depend on socioeconomic conditions. In particular, labor informality, poverty, and socioeconomic strata drive uneven urban mobility changes during the lockdown. Higher socioeconomic strata are consistently associated with higher reductions in mobility. Instead, higher shares of informal workers and a measure of multidimensional poverty are linked to lower decreases in mobility. The general reduction of mobility can be linked to restrictive policies, but self-decisions also drive changes in mobility differently depending on restrictions imposed by socioeconomic conditions. 

\section{Results}

\subsection{Mobility networks}

We build mobility networks counting the number of smart card validations at the integrated public transport system, which includes Trans-Milenio (TM) --a bus rapid transit (BRT) type of transportation system-- and the Integrated Public Transportation System (SITP). We count the number of trips between two stops, defining a trip each time a user makes consecutive validations at two stops. We aggregate trips over working days, and we present mobility data for each calendar week between January 6th and July 17th, 2020. To avoid the effect of holidays, the total number of trips for each week is the average of trips between Tuesday and Thursday. We use the week of February 3rd as the baseline, which represents normal mobility before the lockdown. We analyze mobility ratios that are defined as the mobility flow in a given week over the mobility flow in the reference week (February 3rd) in percentage (see \textit{Materials and Methods} for more details). 
%Seria interesante extenderlo porque el lockdown selectivo genera cambios en la evolucion de la mobilidad

\begin{figure}[h!]
\begin{center}
\includegraphics[width=\textwidth]{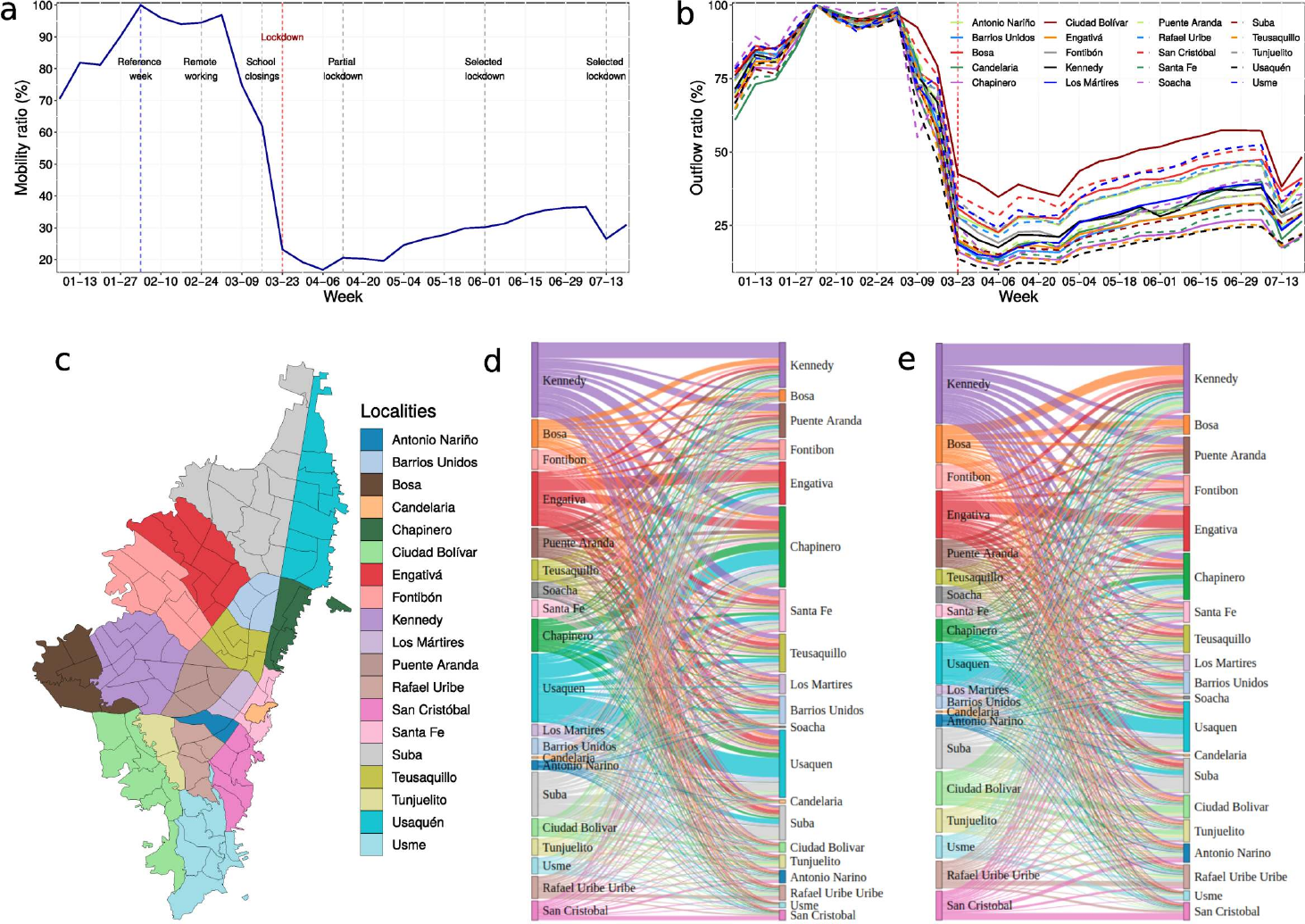}
\caption{\textbf{Evolution of mobility flows in the city of Bogot\'a.} (a) Total mobility ratio in percentage. Vertical lines indicate the implementation of different policies. (b) Outflow ratios (\%) for localities. (c) Map of urban localities and zonal planning units (UPZ). (d-e) Mobility flows between localities in the reference week (February 3rd) and in the lockdown (March 23rd), respectively. Key dates: February 3rd the reference week; February 24th requests of remote working for private companies; March 16th schools closing; March 23rd implementation of strict lockdown; April 14th partial lockdown; June 1st to June 14th, and since July 13th selected strict lockdowns in some localities.}
\label{fig:mobility}
\end{center}
\end{figure}

Both national and sub-national governments in Colombia implemented several measures to contain the spread of COVID-19, prepare the health system, and mitigate the closure measures' economic and social impact. Fig.~\ref{fig:mobility}(a) shows the mobility ratios and the series of measures restricting human mobility in the respective week. We observe that mobility flows are lower in the first weeks of the year because of holidays. We choose February 3rd as the reference week as holidays are over for most of the population, and schools are opened. Although the first case of COVID-19 in Colombia was detected on March 6th, considering other countries' situations, the government issued non-compulsory requests for remote working to private companies already on February 24th, which is associated with a decrease in mobility flows two weeks after the request. Schools, including universities, were closed on March 16th, which is followed by a sharp reduction of mobility flows. The complete lockdown, only allowing essential activities, started on the week of March 23rd, generating a further decrease in mobility flows on that week (23.3\%) and reaching a minimum of 16.8\% two weeks after.

The partial lockdown implementation --since April 14th--, which allowed a gradual restoration of mobility enabling a set of non-essential activities, increases the mobility ratio up to 20.6\% on April 14th. It follows a gradual recovery of mobility flows, reaching a maximum of 36.6\% in the week of July 6th. Since June 1st, some localities started what we name selected lockdown. Although the city was under partial lockdown, these selected localities returned to complete lockdown. In the period considered, this lockdown was implemented for Kennedy, between June 1st and June 14th, and, since July 13th for Santa Fe, M\'artires, Candelaria, Ciudad Bolívar, Rafael Uribe Uribe, and Tunjuelito. The selective lockdown is also associated with a decrease in mobility flows.
%Data on policy implementation are retrieved from \citep{cheng2020covid, hale2020variation}.
%Revisar fecha apertura parcial
%https://www.semana.com/nacion/articulo/coronavirus-en-colombia-el-abc-del-nuevo-decreto-sobre-la-cuarentena-ampliada/662839 

%(excluding the rural locality of Sumapaz)
%-----------------------------
%-----------------------------

Bogot\'a includes 19 urban localities plus one rural, which are subdivided into 117 Zonal Planning Units (UPZ), which are territorial units for planning urban development at the local level and for defining land–use and urban functions (Figure~\ref{fig:mobility}c). Mobility ratios by localities show a sharp decrease (Figure~\ref{fig:mobility}b). However, mobility reduction is faster and reaches a lower minimum level in some localities (for example, Usaqu\'en). Other localities (for example, Ciudad Bol\'ivar) recover their mobility flows faster than others. A similar pattern is observed for mobility inflows by localities. 
Besides, all localities have flows directed to all other localities indicating that the mobility network is highly connected at the locality level, and the connections between localities survive during the lockdown (See: Figs.~\ref{fig:mobility}d-e). We observe changes in the relative relevance of localities, comparing the reference week with the lockdown week. Some localities become less relevant, for example, Teusaquillo, Chapinero, and Usaqu\'en, which are central areas that agglomerate services and workplaces. Instead, other localities gain relative relevance, for example, Kennedy, Ciudad Bol\'ivar, Tunjuelito, and San Crist\'obal, all of them including areas of high urbanization with heterogeneous but, on average, less favourable socioeconomic conditions.

\begin{figure}[h!]
\begin{center}
\includegraphics[width=\textwidth]{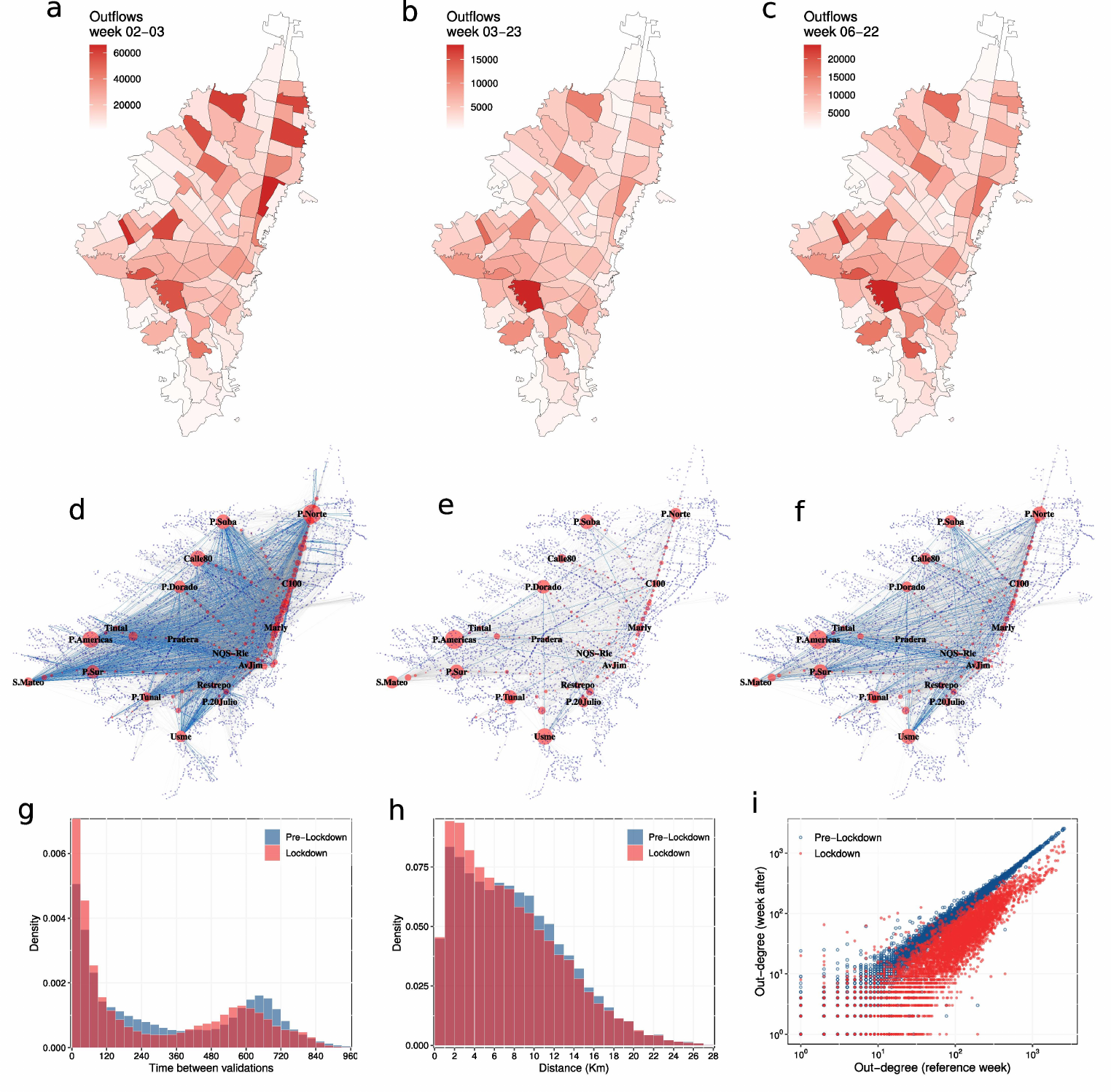}
\caption{\textbf{Changes in mobility flows.} (a-c) Choropleth map of total outflows at UPZ geographical level in different weeks. (d-f) Mobility networks at the stop-level. Nodes are bus stops, and edges are the number of consecutive smart card validations observed during a working day. Nodes in blue belong to the SITP, and in red to TM. Link weights $<$ 10 are filtered out. Link weights $\geq10$ and $<200$ are in grey. Link weights $\geq200$ are in blue. Mobility networks in different weeks: pre-lockdown or reference week (February 3rd), lockdown (March 23rd), and partial lockdown (April 14th), respectively. (g-h) Time and distance distribution, respectively, between smart card validations at the integrated public transport system. Pre-lockdown and lockdown. (i) Correlation between out-degree in the reference week (February 3rd) and out-degree in pre-lockdown (February 10th) and lockdown (March 23rd).}
\label{fig:net_all}
\end{center}
\end{figure}

Of course, changes in mobility are very heterogeneous in more disaggregated geographical areas. The maps in Fig.~\ref{fig:net_all}(a-c) show mobility outflows at the UPZ level in the reference week, in the lockdown, and the partial lockdown. Fig.~\ref{fig:net_all}(d-f) shows mobility networks at the stops-level, before the lockdown (in the reference week, February 3rd), during the lockdown (March 23rd), and during the partial lockdown (June 22nd). %Nodes are stops, and links are trip flows. Red nodes are TM stops, and blue nodes are SITP stops. Nodes' size is determined by the total flows (outflows and inflows), and more relevant nodes include an abbreviated name of the stop. Given that the mobility network consists of 6,197 stops, to improve visualization, link weights between 10 and 200 are presented in grey, while link weights larger than 200 appear in blue.

%-----------------------------
%-----------------------------

The networks reveal the differences between the two types of transportation. The TM shows massive flow of passengers, while the SITP covers a larger geographical area but transports fewer passengers. They also reveal the importance of the economic center of the city, located between the TM stops ``Avenida Jim\'enez" (Av Jim) and ``Calle 100" (C100), with a distance of over 11 km between the stops. This central area receives flows from all the geographical areas of the city. Other relevant nodes in the networks are located in areas of high urbanization. This structure might indicate that population from all areas of Bogot\'a depend on this central area, which agglomerates economic, financial, health, and marketing activities, to access different types of services. 

Comparing mobility networks, we observe that mobility flows drastically decreased in the lockdown week. We observe a substantial reduction in mobility flows, which in the lockdown are 23.2\% of the flows in the reference week (pre-lockdown) (Fig.~\ref{fig:net_all}(a-b)). The gradual reopening of the city leads to a recovery of mobility flows, reaching 35.5\% of the flows observed in the reference week (Fig.~\ref{fig:net_all}(c)).

The mobility network density at the stops-level is low: 2.06\% before the lockdown and 0.78\% in the lockdown, which is not surprising because there are 6,197 stops. Instead, the network's density at the level of UPZ reveals a highly connected network. The density decreases from 90.07\% before the lockdown to 81.21\% after the lockdown. Thus, although mobility flows substantially decreased during the lockdown, the connectivity between different areas of the city remained (see Fig.~\ref{fig:networks}(a-b)).

The distribution of time between consecutive smart card validations for all trips is bimodal, with a large concentration of short-lasting trips (less than 6 hours), and long-lasting trips (between 6 and 16 hours) in the second mode (Fig.~\ref{fig:net_all}(d)). The probability mass around the first mode is right-skewed, while the second one has a clear central tendency between 10 and 11 hours. The shape of the distribution remains in both periods, although we observe a left-shift of approximately one hour. This shift might be explained by less congestion and waiting times during the lockdown.

The distribution of the distance between consecutive validations is right-skewed, and a critical extent of the probability mass concentrates until 14 km (Fig.~\ref{fig:net_all}(e)). This evidence reveals that public transport users in Bogot\'a travel relatively long distances, considering that the city has 1,775 km$^2$, and the distance is more extensive from south to north than from west to east. For example, the distance from the historical center to the biggest station at the north (Portal Norte) is 17.8 km, and to the biggest station in the south (Portal Sur) is 12.5 km. We observe that users in the long-lasting mode typically travel between 4 and 16 km, while those in the short-lasting mode usually travel shorter distances (Fig~\ref{fig:PDF_time}(a-b)).

Before the lockdown, 42\% of the distribution of trips are long-lasting trips, while in the lockdown, the share decreases to 38\%. Although we are not able to determine the reason of these trips, we can assume that the second mode in the distribution of Fig.~\ref{fig:net_all}(d) is mainly capturing workers' trips, considering that the Colombian legal workday has 8 hours and part-time work has 4 hours. The decrease in the share of long-lasting trips might reflect that several employers implemented remote working systems and that mainly workers in essential activities need to travel during the lockdown. On the other hand, before the lockdown, 58\% of the trip flows last less than 6 hours, while after the lockdown, this share reaches 62\% of total flows. A possible explanation for this could be that people in Bogot\'a are constrained to move to access different types of services, including health care, and that the necessity to travel for these reasons persists during the lockdown. %Another possible explanation could be that informal workers make a share of these trips.

The mobility reduction is not only observed in the volume of passengers but also in the number of connections between stops. Fig.~\ref{fig:net_all}(i) shows the changes in the out-degree (number of destinations), for all the stops in a week of relative normality (February 10th), and the week of lockdown (March 23rd), taking as a reference the out-degree in the reference week (February 3rd). Before the lockdown, there is a low dispersion between different weeks. Instead, during the lockdown, the dispersion increases. Thus, although many links remain connected, there is a decrease in the number of destinations for practically all stops.

In sum, the mobility network architecture in different periods reveals interesting features, which can have important implications for policy interventions and urban planning. First, it is highly connected at the UPZ level, indicating the existence of flows between all the city's zones and localities. Secondly, it shows that despite the sharp decrease in mobility flows during the lockdown, the time and distance distributions remain relatively stable, revealing resilience in trip patterns. Thirdly, changes in urban mobility since the lockdown and subsequent recovery show differences at the UPZ level. Probably, differences in socioeconomic conditions might be driving uneven changes in mobility patterns.

\subsection{Changes in mobility and socioeconomic conditions}

We now analyze the relationship between changes in mobility and indicators of socioeconomic conditions of the population. The correlations between mobility outflows and inflows and the share of informal workers, multidimensional poverty, socioeconomic strata (SES), and vulnerability are very weak for the first weeks of the year (Fig.~\ref{fig:corr_evo_inf}(a-b)). Instead, the correlations become significant and positive with informality and poverty, and significant and negative with socioeconomic strata and vulnerability, after the implementation of lockdown policies. 

These correlations indicate that the population with informal jobs travels more since the lockdown compared to formal workers. Similarly, higher levels of poverty are associated with higher mobility flows, while higher socioeconomic strata are associated with lower mobility flows. The population classified as vulnerable drastically reduces mobility during the lockdown. Therefore, mobility restriction measures targeting vulnerable populations (for example, strict lockdown for people over 70 years old and people with comorbidities) and self-decision of the vulnerable population, might also drive mobility changes.

%-----------------------------
\begin{figure}[h!]
\begin{center}
\includegraphics[width=\textwidth]{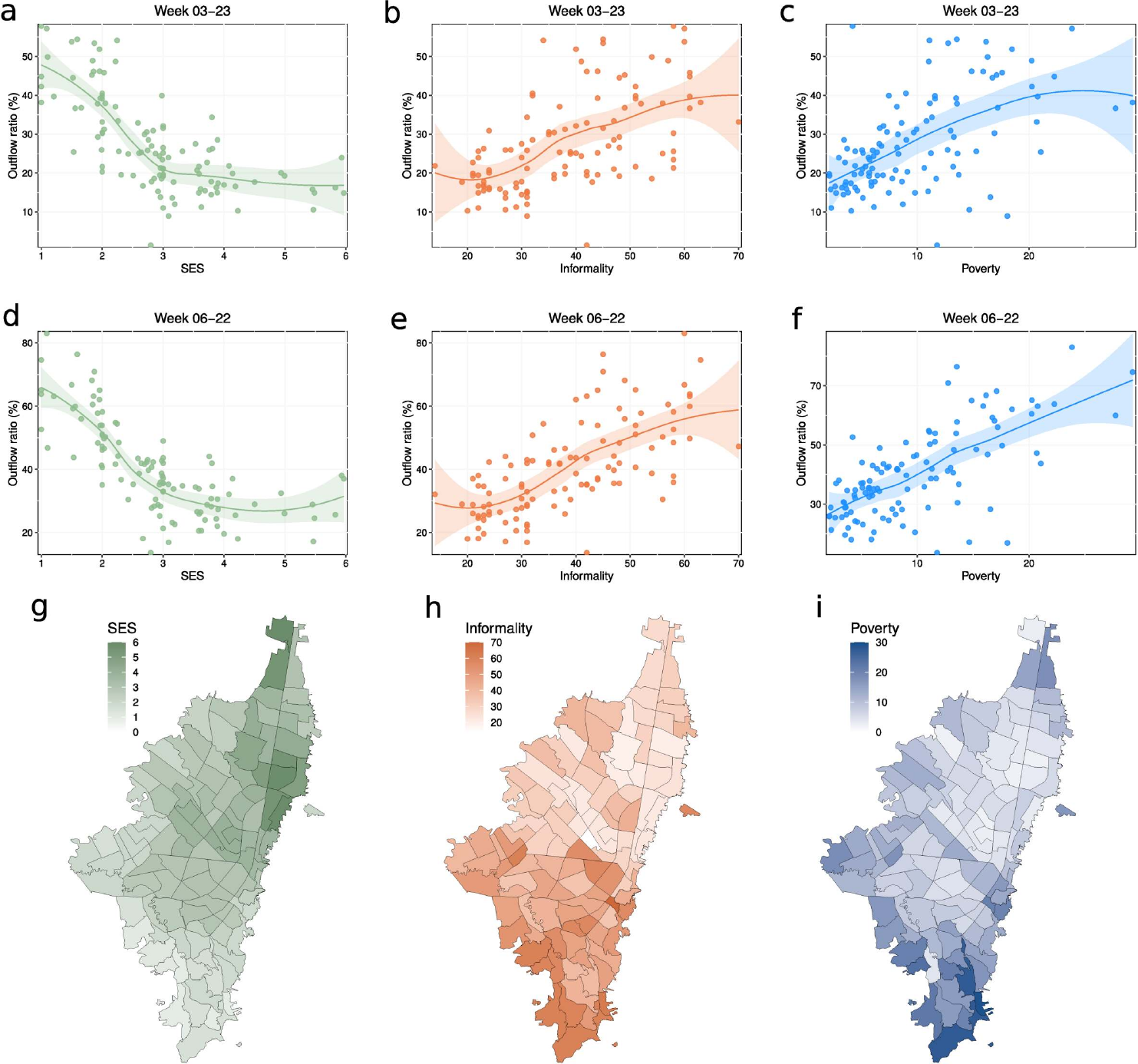}
\caption{\textbf{Changes in mobility and socioeconomic conditions.} Correlations between changes in mobility outflows at the UPZ level with respect to the reference week (February 3rd) and socioeconomic indicators. Shaded areas are 95\% confidence intervals. The vertical dashed red line indicates the beginning of the strict lockdown (March 23rd). (a-f) Correlations between outflows ratio and socioeconomic strata (SES), the share of informal workers, and multidimensional poverty, respectively. (g-i) Socioeconomic conditions at the UPZ level. Maps exclude rural areas.}
\label{fig:socioec}
\end{center}
\end{figure}
%-----------------------------

Figure~\ref{fig:socioec}(a-f) shows the correlation between mobility ratio and socioeconomic strata, the share of informal workers, and multidimensional poverty in the week of the lockdown (March 23rd) and the week of the beginning of the partial lockdown (April 14th). We observe a general decrease of mobility flows. However, the correlations indicate that the reductions depend on the socioeconomic conditions. We observe larger reductions in mobility flows for the population with better socioeconomic conditions and the opposite for worse socioeconomic conditions. During the partial lockdown, the gap between reductions in mobility flows for lower and higher SES, poverty, and share of informal workers, increases, implying that the population with better socioeconomic conditions can maintain a lower mobility for a longer time. 

We build a gravity model to explore the effect of socioeconomic conditions on mobility flows between different zonal planning units, which allows us to address the heterogeneity in socioeconomic conditions (Fig.~\ref{fig:socioec}(g-i)). We estimate the effects of population, distance, SES, informality, and multidimensional poverty, before and after the lockdown, for all trip flows, long-lasting trips, and short-lasting trips (Table~\ref{tb:gm_results} presents the estimation results). 

The results indicate that the geographical distance between the origin and destination has a negative effect on mobility flows, which becomes stronger since the lockdown. Before the lockdown, a higher population in the origin increases mobility. In contrast, the effect of the population in the destination depends on the model and type of trips. In particular, the effect of the population is negative for long-lasting trips, which are assumed to represent formal workers' trips better. The factors attracting flows in the destination are urban amenities, including services and workplaces, which are not necessarily related to the area's population. In Bogot\'a, more urbanized zones are not necessarily those with more and better amenities. Since the lockdown, the effect of the population of the origin and destination is always positive and stronger. Thus, trips are more likely to be observed between more populated zones. 

We analyze the impact of three different indicators of socioeconomic conditions (SES, the share of informal workers, and multidimensional poverty) using independent estimations because these variables are highly correlated (Table~\ref{tb:correlations}). 

Table~\ref{tb:changes} shows the differences in the estimated effects on mobility flows for the lowest and highest scores of the socioeconomic variables at the origin and destination, before and after the lockdown, for the mobility network of all trips, and the networks restricted to long-lasting and short-lasting trips.

%*-*-*-*-*-*-*-*-*-*-*-*-*-*-
\begin{table}[h!]
\begin{center}
\small 
\caption{Relative differences in the estimated mobility flows before and after the implementation of lockdown policies for different socioeconomic conditions at the origin and destination}
\label{tb:changes}
\begin{tabular}{l cc cc cc} 
\toprule
	&	\multicolumn{2}{c}{All trips}		&	\multicolumn{2}{c}{Long-lasting trips}			&	\multicolumn{2}{c}{Short-lasting trips}			\\
%\cmidrule(lr){1-1}
\cmidrule(lr){2-3}
\cmidrule(lr){4-5}
\cmidrule(lr){6-7}	
Variable &	Before	&	After	&	Before 	&	After	&	Before &	After \\
\cmidrule(lr){1-1}
\cmidrule(lr){2-3}
\cmidrule(lr){4-5}
\cmidrule(lr){6-7}	
%\midrule
%	&	\multicolumn{6}{c}{Differences in origins}		\\
%\midrule	
SES$_o$ & -0.35 & 0.54 & -0.08 & 1.51 & -0.53 & 0.05 \\
Informality$_o$ & -0.46 & 0.12 & -0.43 & 0.40 & -0.52 & 0.18 \\
Poverty$_o$ & -0.52 & 0.24 & -0.59 & 0.18 & -0.48 & 0.21 \\
%\midrule
%	&	\multicolumn{6}{c}{Differences in destinations}				\\
\midrule					
SES$_d$ & -0.91 & -0.85 & -0.94 & -0.91 & -0.86 & -0.78 \\
Informality$_d$ & -0.78 & -0.57 & -0.88 & -0.75 & -0.67 & -0.40 \\
Poverty$_d$ & -10.81 & -10.74 & -10.85 & -10.83 & -10.78 & -10.67 \\
\bottomrule
\multicolumn{7}{p{12cm}}{\textit{Note:} The estimated differences are between the SES 1 and 6, the minimum and maximum shares of informal workers observed (14 and 70), and the minimum and maximum levels of poverty observed (2.15 and 29.26). See Table~\ref{tb:summary} and~\ref{tb:gm_results}.}\\
\end{tabular}
\end{center}
\end{table}
%*-*-*-*-*-*-*-*-*-*-*-*-*-*-

Before the lockdown, better socioeconomic conditions at the origin are always positively associated with mobility flows. Considering all trips, the flows of the lowest strata are 35\% lower than those of the sixth socioeconomic strata. Similarly, zones with the highest shares of informal workers have 46\% fewer flows than zones with the lowest share of informal workers, and zones with the highest poverty level generate 52\% fewer flows than the zones with the lowest poverty.

The effects of the socioeconomic conditions in the destinations have more considerable differences than the socioeconomic conditions in the origin. We estimate differences in the flows for extreme values of socioeconomic variables in the destination of -91\% for socioeconomic strata, -78\% for zones with the highest share of informal workers, and ten times less for the highest level of poverty.

%Therefore, the estimations indicate that before the lockdown, zones with better socioeconomic conditions have higher mobility than zones with worse socioeconomic conditions. Instead, although all zones decreased their mobility flows during the lockdown, the reductions are much higher in zones with better socioeconomic conditions. 

We observe that better socioeconomic conditions in the origin lead to a higher reduction of mobility for all trips. We estimate that the flows of the lower socioeconomic strata are 54\% larger than those of the sixth socioeconomic strata. Zones with the highest share of informal workers generate 12\% more flows than zones with the lowest informality, and zones with the highest level of poverty have 24\% more flows than zones with the lowest level of poverty.

During the lockdown, the estimated differences between zones with better and worse socioeconomic conditions slightly decrease. Thus, the socioeconomic conditions of the origin are more relevant to explain changes in mobility flows, which can be related to the strong directionality of the network.

The estimated differences for long- and short-lasting flows are also interesting. In particular, long-lasting trips reveal that the changes in mobility patterns can be larger for the zones with low socioeconomic strata or with higher shares of informal workers. Assuming that mainly formal workers are those explaining long-lasting trips, these larger differences, might derive from differences in the feasibility of working from home.

Overall, the evidence indicates that the population with better socioeconomic conditions have higher possibilities of decreasing mobility. In contrast, a large share of the population with worse socioeconomic conditions needs to travel despite the measures implemented to restrict mobility. Interestingly, we observe a similar behavior at a national level.

\begin{figure}[h!]
\begin{center}
\includegraphics[width=0.9\textwidth]{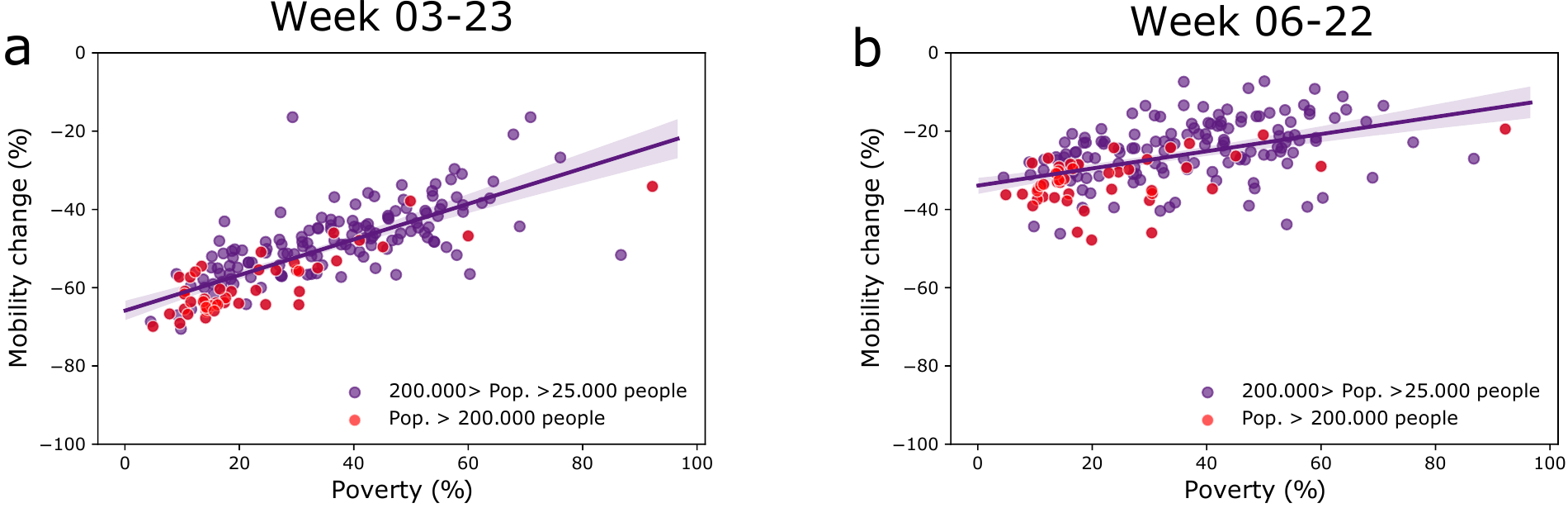}
\caption{\textbf{Mobility changes vs. poverty across the country.} We explore the relationship between the variation in mobility and poverty levels at a municipality level for (a) national lockdown week and (b) the selected week of partial lockdown. We distinguish large urban areas with more than 200.000 inhabitants (red) from medium-size towns with more than 25.000 inhabitants (purple). The correlation is good in both cases, and it seems not to depend on the city's size. In the case of partial lockdown conditions, the high dispersion of points can be due to the different lockdown strategies the local administrations imposed.}
\label{fig:national_level}
\end{center}
\end{figure}

To assess the mobility change at a municipality level, we leverage the Movement Range Maps released by the Facebook – Data for Good program. Specifically, we use the Change in Movement metric that looks at how much people are moving around and compares it to the baseline period, full February 2020 (see \textit{Materials and Methods}). We thus combine these movement data and the multidimensional poverty index aggregated by municipalities. Figure \ref{fig:national_level} shows a clear correlation between these two metrics for the lockdown week and the selected week of partial lockdown. In line with what we observed in Bogot\'a, the richest cities move less than the poorer ones.

%-----------------------------
%-----------------------------

\section{Discussion}

Decreasing urban mobility is key to reduce the spread of COVID-19. Policies implemented to reduce human mobility and self-decisions of citizens, appear effectively driving changes in mobility in the city of Bogot\'a. We observe that, two weeks after the strict lockdown implementation, mobility flows decrease until 16.8\% of the flows observed in the pre-lockdown period, and start gradually recovering during the partial lockdown period until a maximum level of 36.6\% before the application of selected lockdowns. The mobility network is highly connected at the locality and UPZ levels, even during the period of lockdown and partial lockdown. It preserves its main features, revealing the resilience of mobility flows and, in general, of the transportation system.

The average reduction in mobility hides more complex situations related to the ability of the population to respond to social distancing and mobility restriction measures in a middle-income country with high levels of inequality. Changes in mobility are strongly associated with the socioeconomic conditions of the population. Zones with higher poverty, larger shares of informal workers, and lower socioeconomic strata have more moderate mobility reductions during the lockdown than zones with better socioeconomic conditions. Likewise, mobility flows in zones with worse socioeconomic conditions recover faster during the partial lockdown period than zones with better socioeconomic conditions. This evidence reflects that different socioeconomic conditions imply asymmetries in the feasibility of working from home, use savings, postpone consumption, and, more generally, in the ability to respond to lockdown policies positively.

These results have several policy implications. First, if measures restricting mobility are still necessary to contain the spread of COVID-19 in the near future, the evidence indicates that an effective policy of income support needs to be adopted. Otherwise, social distancing will be difficult to achieve because a relevant share of the population will not be able to respond positively to these measures. In the short-term, an income support policy could help this population, but improving general living conditions and reducing inequality will be necessary in the medium-term. 

Secondly, short-lasting trips show a lower reduction, indicating that not only workers in essential and allowed activities are constrained to move but also other people who probably move to access services, including health care, employment, recreation, and food. A more detailed analysis uncovering the reasons for these trips is necessary. However, different measures could be implemented to reduce this type of flows. For example, decentralize legal and healthcare systems, grant universal internet access, and improve technology platforms to provide online public and private services. In sum, implement policies aiming to redistribute activities so that people can access them without using public transport, this is, by bicycle or on foot.
%system congestion must be mitigated
%passenger limits in public transport

%escalonamiento en horarios de trabajo segun actividades
%entender tipo de actividades q se desarrollan en las upz (oferta y demanda de actividades) permitiria proponer movimientos escalonados

Thirdly, the transport system in Bogot\'a has several problems. The mobility network's resilience indicates that these problems persist during the lockdown and will return as soon as the system recovers mobility flows. Therefore, addressing the transport system's problems is imperative in the context of a gradual reopening of economic activities and mobility. Policymakers and universities are proposing or evaluating possible measures to address these problems. For example, generate alternative transport routes, redistribute the programming of routes to dispatch more buses to areas of high urbanization and worse socioeconomic conditions, promote the staggering of working hours to reduce congestion, guarantee additional resources to increase frequencies and changes in operation, adopt disinfection practices in stations and buses, facilitate the use of the bicycle providing a connected and well designed infrastructure, and generate changes in social behavior (see, for example: \citep{medidas}). 
%1. Generate alternative transport routes to the Transmilenio System to guarantee the service to the citizens that must be mobilized during the quarantine.
%2. Redistribute the programming of the routes to dispatch more buses to the orange and red stations (see map) concentrated in the south of the city.
%3. Promote the staggering of working hours and thus reduce congestion at the entrance to stations with the greatest number of users.
%4. Encourage physical distancing and the use of masks in those who use the system.
%5. Facilitate the operation so that it is possible for all the people on a bus to be seated with a chair between the other occupants.
%6. Request an emergency operation subsidy from the national government during quarantine to cover the costs of increasing frequencies and changes in operation.
%7. Recommend and facilitate the conditions for the use of the bicycle for those who can do it.

The impact of human mobility and travel restrictions on the spread of COVID-19 has been largely reported in several studies so far \citep{kraemer2020effect, gatto2020spread, hadjidemetriou2020impact}. While intra-city travel reductions have a high impact on overall infection numbers, intercity travel reductions can narrow the scope of the outbreak and help target resources. The empirical evidence displayed in this paper shows that high poverty levels in middle-income countries drive to non-compliance with lockdown policies and social-distancing. These non-compliance behaviors could prevent the fast and successful containment of the pandemic, as partly observed in Latin America. Future research needs to systematically assess the contributions of these factors in the evolution of disease transmissions and, thus, in how to reduce their impact.

Overall, in a city with significant inequalities, the needs and possibilities of the vulnerable population due to socioeconomic conditions, must be addressed to implement successful strategies of social distance and mobility restriction. 

\section{Material and methods}

\subsection{Geographical units}

Bogot\'a consists of 19 urban localities plus one rural, which are subdivided into 117 Zonal Planning Units (UPZ). Social and economic conditions are very heterogeneous in Bogot\'a, and this heterogeneity persists even within UPZ. Despite this, we use the UPZ as our main geographical unit of analysis since, compared to localities, it allows better control of social and economic heterogeneity. Moreover, UPZ are units of interest in urban planning.

\subsection{Socioeconomic data}
Several indicators of socioeconomic conditions in Bogot\'a derive from geolocated Census data with a high spatial resolution. We use indicators available at the block level and aggregate them at the UPZ level, weighting variables of interest by the population of each UPZ. 
%which are available at \citep{DANE_visor}, and the indicator of socioeconomic strata.

\subsubsection*{Population and socioeconomic strata}
%Population data are from the 2018 National Population and Housing Census and the Individual Registry of Health Benefits.
We obtain a census population dataset at the city block-level with their socioeconomic strata (SES). There are around 45,000 blocks classified by SES in the city. This classification has been made according to external conditions and urban surroundings, including the quality of urban spaces and access to goods and services, with stratum from 1 to 6, representing the resident's income level from low to high \citep{uribe2008estratificacion}. The socioeconomic stratification targets houses instead of families or individuals, and it is defined by a formal body of the Stratification Committee \citep{gallego2016effect}. The strata 1, 2, and 3, include 86,04\% of the population, the strata 4 includes 9,42\%, and the strata 5 and 6 include only 4,54\% of total population \citep{secretaria2017}. This dataset is used for generating population-weighted SES estimates at the UPZ level. 

\subsubsection*{Multidimensional poverty}
Multidimensional poverty has five dimensions (educational conditions, conditions of childhood and youth, health, work, access to public services, and housing conditions), and 15 indicators. All dimensions weight 20\%, and the indicators have the same weight within their respective dimension. Households are considered poor if they have deprivation in at least 33.3\% of the indicators \citep{DANE_poverty}.

\subsubsection*{Vulnerability}
The indicator of vulnerability is built by determining the share of individuals per block with a set of comorbidities, which are identified as risk factors that can generate complications in people with COVID-19, the share of people over 60 years old and in an overcrowded household or living alone, and population density per block. The indicator is built using the 2018 National Population and Housing Census and the Individual Registry of Health Benefits \citep{DANE_vulnera}.

%\subsubsection*{Socioeconomic strata}
%The city of Bogot\'a is classified by socioeconomic strata (SES), according to external conditions and urban surroundings, including the quality of urban spaces and access to goods and services, with stratum from 1 to 6, representing the resident's income level from low to high \citep{uribe2008estratificacion}. The socioeconomic stratification targets houses instead of families or individuals, and it is defined by a formal body of the Stratification Committee \citep{gallego2016effect}. The strata 1, 2, and 3, include 86,04\% of the population, the strata 4 includes 9,42\%, and only 4,54\% of total population is in the strata 5 and 6 \citep{secretaria2017}.

\subsubsection*{Informality}
 As an effort to characterize informal employment in the city, the District Department of Planning in Bogot\'a (SDP) has estimated the informality at UPZ level~\cite{DianaGutierrez}. Defined as the share of individuals having an occupation with an income but not enrolled in the health system or pension fund~\citep{informal}, we used this dataset for analyzing its correlation with the mobility reduction in the same areas.
 %Informality measures the number of individuals with informal jobs for each locality of the city. It is defined as the share of individuals having an occupation with an income but not enrolled in the health system or pension fund \citep{informal}.

%Policy

%Data on policy implementation is retrieved from \citep{cheng2020covid, hale2020variation}.

%An index that quantifies government responses to COVID--19 \citep{hale2020variation}.

\subsection{Mobility data}

\subsubsection*{Public transport data}
We use public transport data from the city of Bogot\'a. Public bus transportation in Bogot\'a consists of Trans-Milenio (TM), which is a bus rapid transit (BRT) type of transportation system, and the Integrated Public Transportation System (its acronym in Spanish SITP). According to the Bogot\'a's transportation company, TM has 143 stations, a coverage of 114.4 km, and 98 bus routes, while the SITP has 7,516 stops, a coverage of 1,890.4 km and 282 bus routes \citep{TMreport_feb20}.

To access the transport service, each user must validate its smart card --\textit{Tullave} (which is personal)--, at the entrance of each station in TM or a bus in SITP. We use smart card validations at the stop or station of the integrated public transport system, the date, and the time. Each smart card contains an anonymous code in our data, which allows us to reconstruct users' daily movements.

%More than 2 million validations per day. 
The data have some limitations. We do not have information on exits, which will provide the complete trip of each user. However, given that we have all smart card validations for each user in a day, we assume that the user's full trip is between the first and last validation of a day. Of course, people may use other means of transport (such as taxis, which are commonly employed in Bogot\'a) on the same day. However, for most users, the different validations in a day will provide complete enough indication of its mobility. 

Our mobility data do not provide the city's total flows in a day because we do not consider other transportation means, such as taxis and cars. However, given the volume of users in the transport system, the available data are likely to represent the mobility patterns of Bogot\'a.

We implement a data cleaning procedure, excluding a small percentage of smart cards with a single validation in the day, consecutive validations at the same stop, validations that appear at the bus garages, and trips with more than 10 smart card validations in the same day, which could be related to workers of the transport system. Eliminating these observations, we end up with around 95\% of the validations of a typical workday. Also, we consider all 143 TM stations in the analysis, and we exclude a number of SITP stops that are not geolocated. We end up with a total of 6,197 stops.

\subsubsection*{Facebook movement range data}

Facebook – Data for Good initiative is a collection of methods for processing Facebook data into dynamic spatial-temporal maps that illustrate how populations prepare and respond to a crisis. Among numerous datasets, Facebook made publicly available the Movement Range datasets. These datasets have two different metrics: Change in movement and Stay Put. The change in movement metric reflects the average number of level 16 bing tiles ($0.6$km $\times$ $0.6$ km) that a Facebook user (who enabled his geo-positioning) was present during a 24-hour period relative to the pre-crisis levels. The Stay Put metric measures the fraction of the population that appears to stay within a small area surrounding their home for an entire day. Thus, a stationary user is analogous to staying put or staying home. In the case of Colombia, the data is available at the municipality scale. For this paper, we use the change in movement metric as a proxy of the reduction in mobility.

\subsection{Mobility networks}
To build the mobility network for each week, we consider Tuesdays, Wednesdays, and Thursdays. We avoid using weekends because our primary interest is in the movement on working days. Mondays and Fridays are omitted since the period includes ten holidays, all on Mondays and Fridays, except one on Thursday, April 9th. Also, we exclude Fridays because mobility patterns are likely to shift as it is common to observe flows towards meeting and social recreation places outside the city.

Thus, the Bogot\'a public transport network in a given week ($t$) is represented by a weighted and directed graph, where nodes correspond to all stops (in both SITP and TM), and the link weight is the number of people that make consecutive validations between a given couple of stops. Therefore, the first validation stop corresponds to the origin $o$ and the successive validation stop to the destination $d$.

We also aggregate mobility data at the UPZ level and represent these flows as a network. To do this, we group all stops by their corresponding UPZ and compute the mobility flows to all other UPZ. The network aggregated to this geographical level allows us to detect the relationships between mobility patterns and other local socioeconomic conditions.

\subsubsection*{Network statistics}

We use node statistics which allow studying nodes' characteristics in terms of connectivity for each week. In general, for a generic weighted network matrix $W$ with its corresponding adjacency matrix $A$, the network statistics for node $i$ are: (i) node out-strength $NS_{i}^{out}=W_{(i)}\textbf{1}$, which measures the total outflows; (ii) node in-strength $NS_{i}^{in}=W_{(i)}^T\textbf{1}$, which measures the total inflows; (iii) node out-degree $ND_{i}^{out}=A_{(i)}\textbf{1}$, which measures the number of destinations; and, (iv) node in-degree $ND_{i}^{in}=A_{(i)}^T\textbf{1}$, which measures the number of origins connecting node $i$; where \textbf{1} is a vector with ones in all entries; and, $A_{(i)}$ and $W_{(i)}$ are the $i$ rows of $A$ and $W$, correspondingly. Network density is the ratio between the actual number of connections and the potential connections: $\rho=m/n(n-1)$, where $m$ is the number of links and $n$ is the number of nodes. 

\subsection{Gravity model estimations}
Let $W_{od}(t)$ be the mobility flow between origin $o$ and destination $d$, in week $t$. We estimate the following gravity equation:
\begin{equation}\label{eq:gravity}
W_{od}(t) = \frac{Pop_o^{\alpha_1} Pop_d^{\alpha_3}} {D_{od}^{\alpha_3}} \exp\{x_{od}(t) \cdot \beta\} \eta_{od}(t),
\end{equation}
where,
\begin{equation}\label{eq:var}
x_{od}(t)=\{X_{o}, X_{d}, \tau_t\};
\end{equation} 
$o,d=1,...,N$; $Pop_o$ is the population at location $o$ and $Pop_d$ is the population at location $d$; $D_{od}$ is the geographical distance between locations $o$ and $d$; $X=\{$SES, Informality, Poverty$\}$ is a vector of socioeconomic indicators, including the socioeconomic strata, the share of informal workers, and the level of multidimensional poverty, of the location; $\tau_t$ is a set of weekly time dummies; and we assume that $E[\eta_{od}|Pop_o,Pop_d,D_{od},...]=1$. 

There are different econometric techniques to estimate Eq.~\eqref{eq:gravity}. Simple ordinary least squares (OLS) could be applied to the log-linearized equation. However, the estimation requires special treatment of heteroskedasticity (non-linearity) and zero-valued flows \citep{santos_tenreyro_2006}. Thus, to deal with these methodological challenges, we implement a Poisson estimation method, which has shown quite satisfactory goodness of fit compared to other econometric techniques to describe network's properties \citep{duenas2013modeling, duenas2014global}.

\bibliographystyle{vancouver}
\bibliography{biblio}

\newpage
\section*{Supplementary Information}\label{sec:A}
\setcounter{table}{0}\renewcommand{\thetable}{SI.\arabic{table}}
\setcounter{figure}{0}\renewcommand{\thefigure}{SI.\arabic{figure}}

\subsection*{\textbf{Changes in mobility and socioeconomic conditions in Bogot\'a city during the COVID-19 outbreak}}

\vspace{1cm}

\subsection*{Network statistics}

\begin{figure}[h!]
\begin{center}
\includegraphics[width=\textwidth]{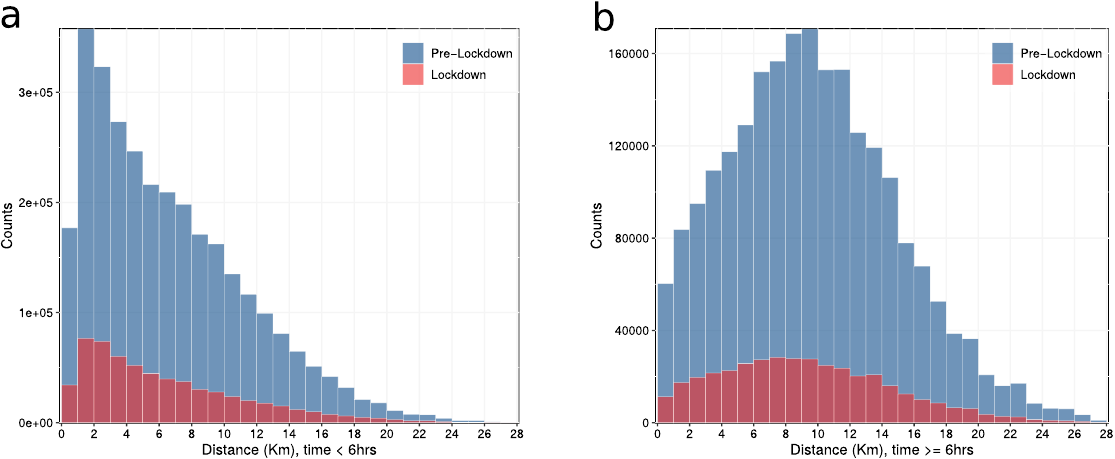}
%\includegraphics[width=0.35\textwidth]{PDF_dist_part_time.pdf}
%\hspace{10pt}
%\includegraphics[width=0.35\textwidth]{PDF_dist_full_time.pdf}
\caption{\textbf{Distribution of time between smart card validations.} (a) Short-lasting trips ($<$ 6 hours). (b) Long-lasting trips ($\geq 6$ hours). (a-b) Pre-lockdown (February 3rd) and lockdown (March 23rd).}
\label{fig:PDF_time}
\end{center}
\end{figure}

Fig.~\ref{fig:PDF_time} shows the distribution of time between smart card validations at the integrated public transport system for trips lasting less and more than 6 hours, respectively, for the pre-lockdown and the lockdown periods.

%-----------------------------
%-----------------------------

\begin{figure}[h!]
\begin{center}
\includegraphics[width=\textwidth]{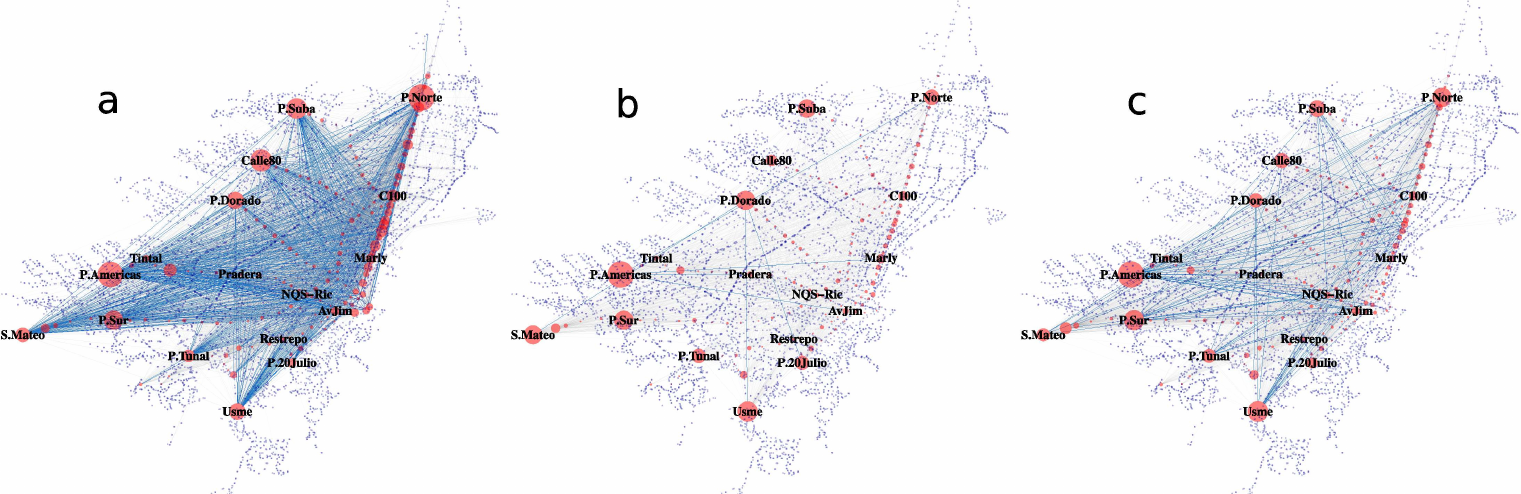}
%{\includegraphics[width=0.328\textwidth]{Net_Workers_prelock.pdf}}
%{\includegraphics[width=0.328\textwidth]{Net_Workers_lock.pdf}}
%{\includegraphics[width=0.328\textwidth]{Net_Workers_postlock.pdf}}
\caption{\textbf{Mobility networks of long-lasting trips ($\geq 6$ hours).} (a) Pre-lockdown (February 3rd). (b) Lockdown (March 23rd). (c) Partial lockdown (April 14th). (a-c) Nodes are bus stops and edges are the number of consecutive smart card validations observed during a working day. Nodes in blue belong to the SITP, and in red to TM. Link weight $<$ 10 filtered out. Link weights $\geq10$ and $<200$ are presented in grey. Link weights $\geq200$ are presented in blue.}
\label{fig:net_long}
\end{center}
\end{figure}

\begin{figure}[h!]
\begin{center}
\includegraphics[width=\textwidth]{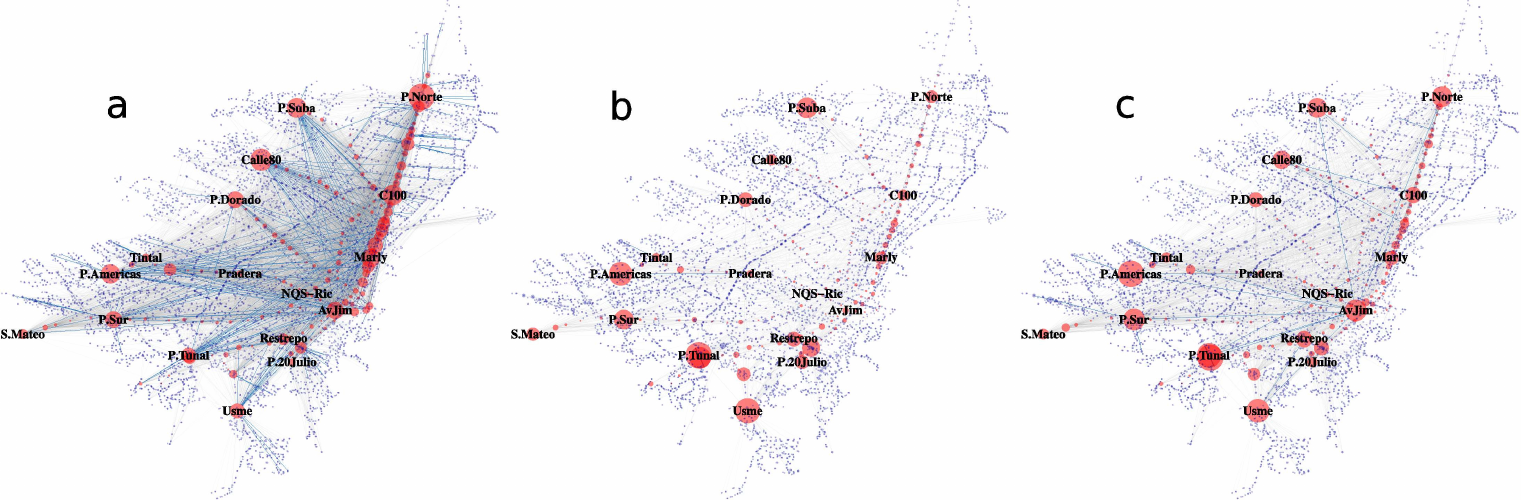}
%{\includegraphics[width=0.328\textwidth]{Net_PartTime_prelock.pdf}}
%{\includegraphics[width=0.328\textwidth]{Net_PartTime_lock.pdf}}
%{\includegraphics[width=0.328\textwidth]{Net_PartTime_postlock.pdf}}
\caption{\textbf{Mobility networks of short-lasting trips ($<$ 6 hours).} (a) Pre-lockdown (February 3rd). (b) Lockdown (March 23rd). (c) Partial lockdown (April 14th). (a-c) Nodes are bus stops and edges are the number of consecutive smart card validations observed during a working day. Nodes in blue belong to the SITP, and in red to TM. Link weight $<$ 10 filtered out. Link weights $\geq10$ and $<200$ are presented in grey. Link weights $\geq200$ are presented in blue.}
\label{fig:net_short}
\end{center}
\end{figure}

Fig.~\ref{fig:net_long} and~\ref{fig:net_short} show the mobility networks of long-lasting and short-lasting trips, respectively, for the reference week, before the lockdown, the lockdown week, and the selected week of partial lockdown. 

%-----------------------------
%-----------------------------

%-----------------------------
%-----------------------------

%En la red de short: los flujos que mas se reuperan son los que van de las zonas pobres a otras, y esto nos permite asumir que el motivo es que necesitan viajar paa acceder a servicios. 

\newpage
To detect changes in mobility patterns in the city of Bogot\'a, we analyze the evolution of the network density and the correlation structure of nodes' characteristics, in particular, the correlations between the node's in-degree and out-degree and the node's in-strength and out-strength. The former considers the relationship between the number of arrivals and departures from a given location, while the latter measures the relationship between the volume of inflows and outflows. These correlations can help detect whether mobility restricting policies have caused changes in mobility patterns at the level of the connection nodes. 

%-----------------------------
\begin{figure}[h!]
\begin{center}
\includegraphics[width=\textwidth]{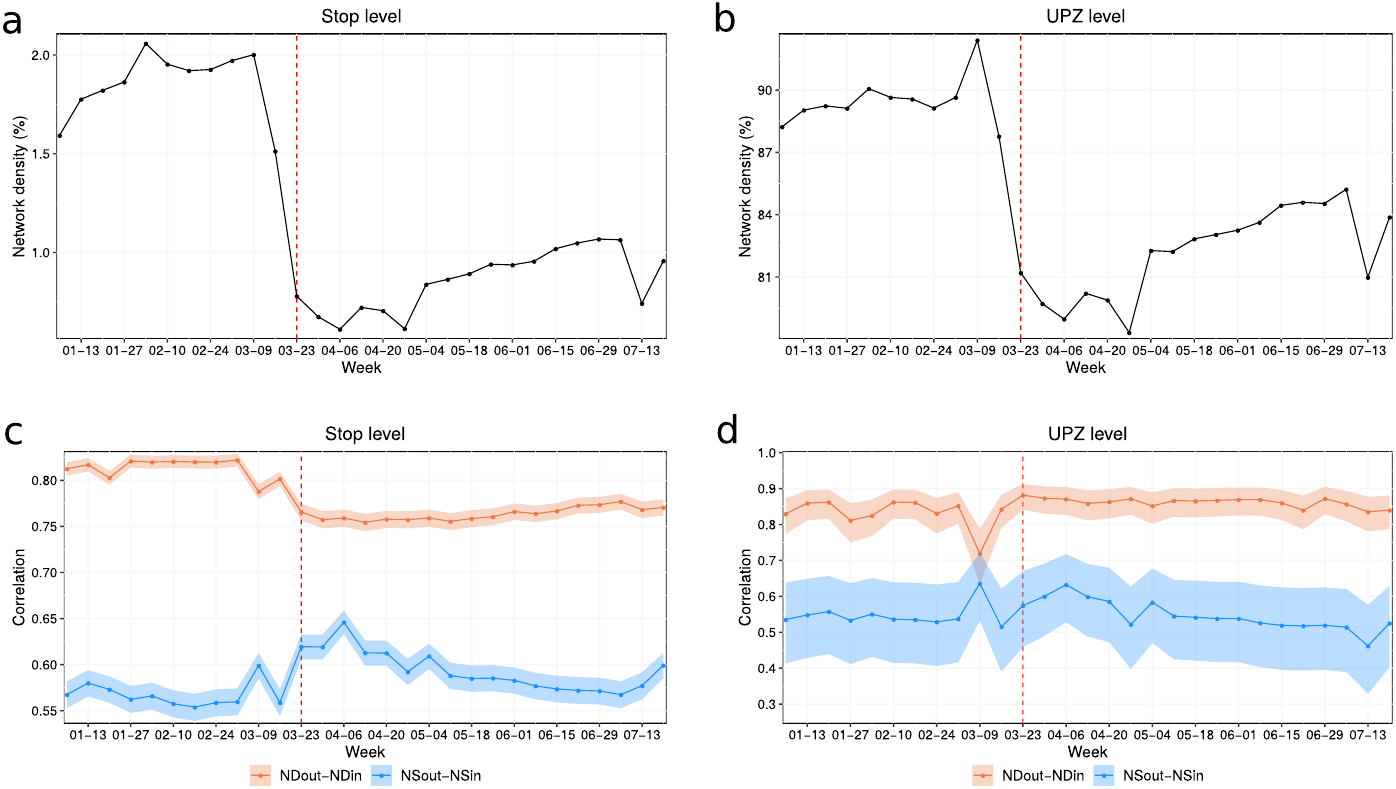}
\caption{\textbf{Evolution of network statistics.} Vertical dashed red lines indicate the beginning of the lockdown (March 23rd). (a-b) Evolution of the network density at the level of stops and at the level of UPZ, respectively. (c-d) Correlation between node's in-degree and out-degree, and between node's in-strength and out-strength, at the stop and UPZ levels, respectively.}
\label{fig:networks}
\end{center}
\end{figure}
%-----------------------------

%Fig.~\ref{fig:networks} shows the evolution of the network density and the correlations for the stops and UPZ levels. 
Not surprisingly, the network's density is very sensitive to the implementation of lockdown policies, especially for the network where nodes are stops, given that there are 6,197 stops. Between the reference week and the lockdown week, the network's density at the UPZ level drops from 90.07\% to 81.21\%, while the density of the network at the stops-level drops from 2.06\% to 0.78\% (Fig.~\ref{fig:networks}(a-b)).

The correlations in Fig.~\ref{fig:networks}(c-d) also change. However, they remain strong and significant, revealing signs of the resilience of mobility patterns. The correlation between the number of arrivals and departures (in-degree and out-degree) at the stops-level drops, revealing higher dispersion among these node statistics than the weeks before the lockdown. Conversely, the correlation between outflows and inflows (in-strength and out-strength) increases. During the lockdown, policymakers preferred to keep active the bus rapid transit system (TM), which can explain the changes in the correlations. Also, the SITP was more affected since its operations are distributed more evenly throughout the city. Interestingly, these correlations seem to move towards the levels they had before the lockdown policies.

The behavior of these correlations at the UPZ level shows no significant changes comparing the weeks before and after the lockdown. This evolution adds information on the behavior of groups of people with similar socioeconomic conditions. The evidence reveals that structurally there are no significant changes in the mobility patterns between UPZ. Even though the effects at the stops-level are more evident, the aggregation at the UPZ level reveals that there are no dramatic changes in the number of arrival and departure destinations or the inflow and outflow volumes. 

\subsection*{Socioeconomic conditions and mobility flows}

\begin{figure}[h!]
\begin{center}
\includegraphics[width=\textwidth]{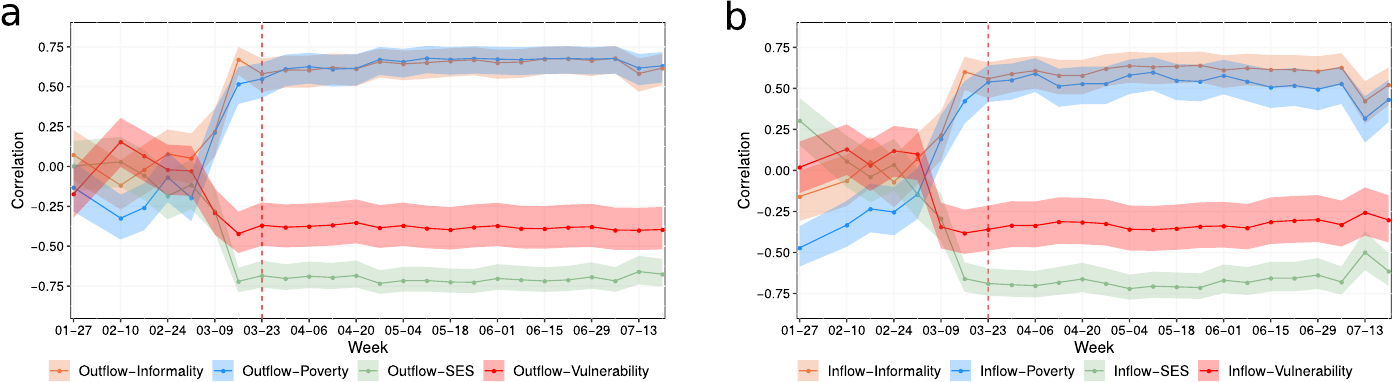}
%\includegraphics[width=0.44\textwidth]{cor_AD_NSout.pdf}
%\hspace{10pt}
%\includegraphics[width=0.44\textwidth]{cor_AD_NSin.pdf}
\caption{\textbf{Correlations between changes in mobility and socioeconomic indicators.} (a-b) Mobility outflows and inflows, respectively, at the UPZ level, with respect to the reference week (February 3rd). Shaded areas are 95\% confidence intervals. Vertical dashed red line indicates the beginning of the lockdown (March 23rd).} 
\label{fig:corr_evo_inf}
\end{center}
\end{figure}

Fig.~\ref{fig:corr_evo_inf} shows the evolution of the correlations between mobility outflows and inflows and socioeconomic conditions. 

%-----------------------------
%-----------------------------

%-----------------------------
%\begin{figure}[h!]
%\begin{center}
%\includegraphics[width=\textwidth]{figSI6.pdf}
%\includegraphics[width=0.3\textwidth]{cor_poverty_nat_12.pdf}
%\hspace{10pt}
%\includegraphics[width=0.3\textwidth]{cor_poverty_nat_20.pdf}
%\hspace{10pt}
%\includegraphics[width=0.3\textwidth]{cor_poverty_nat_30.pdf}
%\caption{\textbf{Changes in mobility flows at the national level and poverty.} (a-c) Correlations between mobility flows at the national level and poverty in different weeks.}
%\label{fig:poverty_nat}
%\end{center}
%\end{figure}
%-----------------------------

%\newpage
\subsection*{Gravity model estimations of bilateral flows}

Table~\ref{tb:correlations} presents the correlations between the indicators of socioeconomic conditions. Not surprisingly, multidimensional poverty, informality, and socioeconomic strata are highly correlated. Table~\ref{tb:summary} presents the summary statistics of the variables used in the analysis. 

%*-*-*-*-*-*-*-*-*-*-*-*-*-*-
\begin{table}[h!]
\begin{center}
\scriptsize 
\caption{Correlation between socioeconomic variables}
\label{tb:correlations}
\renewcommand{\arraystretch}{1.1}
%\begin{tabular}{l C{1.7cm} c C{3cm} C{2.9cm}} 
\begin{tabular}{l ccccc} 
\toprule
 & $\ln$ Population & Vulnerability & MPI & SES & Informality \\
\cmidrule(lr){2-6}
$\ln$ Population & 1 &  &  &  &  \\
Vulnerability & 0.279* & 1 &  &  &  \\
Multidimensional Poverty Index (MPI) & 0.022 & -0.507* & 1 &  &  \\
Socioeconomic strata (SES) & -0.319* & 0.436* & -0.698* & 1 &  \\
Informality & 0.166* & -0.387* & 0.684* & -0.733* & 1 \\
\bottomrule
\multicolumn{6}{p{12cm}}{\textit{Note:} Significance level of 95\% indicated with an asterisk (*).}\\
\end{tabular}
\end{center}
\end{table}
%*-*-*-*-*-*-*-*-*-*-*-*-*-*-

%*-*-*-*-*-*-*-*-*-*-*-*-*-*-
\begin{table}[h!]
\begin{center}
\scriptsize 
\caption{Summary statistics of socioeconomic variables}
\label{tb:summary}
\renewcommand{\arraystretch}{1.1}
%\begin{tabular}{l C{1.7cm} c C{3cm} C{2.9cm}} 
\begin{tabular}{l cc cc cc} 
\toprule
Variable & Obs & Mean & Std. Dev. & Min & Max \\
\midrule
Population & 111 & 78,268 & 65,068 & 179 & 353,991 \\
Vulnerability & 111 & 2.816 & 0.356 & 2.000 & 3.827 \\
Multidimensional poverty & 111 & 9.794 & 5.881 & 2.148 & 29.262 \\
Socioeconomic strata & 111 & 2.890 & 1.108 & 1.000 & 5.966 \\
Informality & 111 & 38.450 & 12.818 & 14.000 & 70.000 \\
\bottomrule
\end{tabular}
\end{center}
\end{table}
%*-*-*-*-*-*-*-*-*-*-*-*-*-*-

Table~\ref{tb:gm_results} shows the estimation results. Models 1-2 considers all trips, models 3-4 restricts the estimation to long-lasting trips, and models 5-6 considers short-lasting trips only. We consider the effect of socioeconomic conditions before and after the lockdown.

%*-*-*-*-*-*-*-*-*-*-*-*-*-*-
\begin{table}[h!]
\begin{center}
\scriptsize
\caption{Gravity model estimations}
\label{tb:gm_results}
\renewcommand{\arraystretch}{1.1}
\resizebox{\textwidth}{!}{\begin{tabular}{l ccc ccc ccc} 
\toprule
Model  & (1) & (2) & (3) & (4) & (5) & (6) & (7) & (8) & (9) \\
\cmidrule(lr){1-1}
\cmidrule(lr){2-4}
\cmidrule(lr){5-7}
\cmidrule(lr){8-10}
 Sample & \multicolumn{3}{c}{All trips}  & \multicolumn{3}{c}{Long-lasting trips} & \multicolumn{3}{c}{Short-lasting trips} \\
\cmidrule(lr){1-1}
\cmidrule(lr){2-4}
\cmidrule(lr){5-7}
\cmidrule(lr){8-10}
$\ln$ Distance$_{od}$  & -0.813*** & -0.810*** & -0.767*** & -0.503*** & -0.448*** & -0.390*** & -1.040*** & -1.059*** & -1.024*** \\
 & (0.008) & (0.009) & (0.009) & (0.009) & (0.010) & (0.010) & (0.009) & (0.008) & (0.009) \\
 $\ln$ distance$_{od} \times$ lockdown  & -0.087*** & -0.073*** & -0.096*** & -0.074*** & -0.069*** & -0.083*** & -0.069*** & -0.050*** & -0.073*** \\
 & (0.010) & (0.011) & (0.011) & (0.010) & (0.012) & (0.013) & (0.011) & (0.010) & (0.011) \\
$\ln$ Population$_o$  & 0.255*** & 0.244*** & 0.221*** & 0.305*** & 0.316*** & 0.298*** & 0.230*** & 0.201*** & 0.174*** \\
  & (0.008) & (0.008) & (0.007) & (0.012) & (0.011) & (0.010) & (0.007) & (0.006) & (0.006) \\
$\ln$ Population$_d$  & 0.074*** & -0.018*** & -0.057*** & 0.025*** & -0.088*** & -0.137*** & 0.133*** & 0.054*** & 0.024*** \\
 & (0.005) & (0.004) & (0.004) & (0.006) & (0.005) & (0.004) & (0.005) & (0.005) & (0.005) \\
 $\ln$ Population$_d \times$ lockdown  & 0.071*** & 0.092*** & 0.119*** & 0.046*** & 0.070*** & 0.099*** & 0.092*** & 0.109*** & 0.135*** \\
  & (0.010) & (0.009) & (0.009) & (0.014) & (0.013) & (0.012) & (0.008) & (0.008) & (0.008) \\
$\ln$ Population$_o \times$ lockdown  & 0.120*** & 0.119*** & 0.142*** & 0.084*** & 0.091*** & 0.111*** & 0.138*** & 0.131*** & 0.155*** \\
 & (0.007) & (0.006) & (0.006) & (0.008) & (0.006) & (0.006) & (0.007) & (0.006) & (0.006) \\
Socioeconomic strata$_o$  & 0.086*** &  &  & 0.016* &  &  & 0.152*** &  &  \\
  & (0.007) &  &  & (0.009) &  &  & (0.007) &  &  \\
Socioeconomic strata$_d$  & 0.474*** &  &  & 0.570*** &  &  & 0.391*** &  &  \\
 & (0.008) &  &  & (0.009) &  &  & (0.007) &  &  \\
Socioeconomic strata$_o \times$ lockdown  & -0.173*** &  &  & -0.200*** &  &  & -0.161*** &  &  \\
  & (0.009) &  &  & (0.010) &  &  & (0.009) &  &  \\
Socioeconomic strata$_d \times$ lockdown  & -0.088*** &  &  & -0.082*** &  &  & -0.086*** &  &  \\
 & (0.009) &  &  & (0.011) &  &  & (0.008) &  &  \\
Informality$_o$  &  & -0.011*** &  &  & -0.010*** &  &  & -0.013*** &  \\
  &  & (0.001) &  &  & (0.001) &  &  & (0.001) &  \\
Informality$_d$  &  & -0.027*** &  &  & -0.038*** &  &  & -0.020*** &  \\
 &  & (0.001) &  &  & (0.001) &  &  & (0.001) &  \\
 Informality$_o \times$ lockdown  &  & 0.013*** &  &  & 0.016*** &  &  & 0.012*** &  \\
  &  & (0.001) &  &  & (0.001) &  &  & (0.001) &  \\
Informality$_d \times$ lockdown  &  & 0.012*** &  &  & 0.013*** &  &  & 0.011*** &  \\
 &  & (0.001) &  &  & (0.001) &  &  & (0.001) &  \\
Poverty$_o$  &  &  & -0.027*** &  &  & -0.033*** &  &  & -0.024*** \\
  &  &  & (0.001) &  &  & (0.002) &  &  & (0.001) \\
Poverty$_d$  &  &  & -0.061*** &  &  & -0.070*** &  &  & -0.056*** \\
 &  &  & (0.001) &  &  & (0.002) &  &  & (0.001) \\
Poverty$_o \times$ lockdown  &  &  & 0.035*** &  &  & 0.039*** &  &  & 0.031*** \\
  &  &  & (0.002) &  &  & (0.002) &  &  & (0.002) \\
Poverty$_d \times$ lockdown  &  &  & 0.011*** &  &  & 0.004* &  &  & 0.015*** \\
 &  &  & (0.002) &  &  & (0.002) &  &  & (0.001) \\
Constant & 1.102*** & 5.358*** & 5.325*** & -0.561*** & 4.033*** & 3.828*** & 0.770*** & 4.874*** & 4.903*** \\
 & (0.125) & (0.091) & (0.090) & (0.172) & (0.119) & (0.118) & (0.110) & (0.086) & (0.086) \\
Observations & 354,090 & 354,090 & 354,090 & 354,090 & 354,090 & 354,090 & 354,090 & 354,090 & 354,090 \\
Pseudo R2 & 0.356 & 0.307 & 0.308 & 0.300 & 0.243 & 0.233 & 0.410 & 0.373 & 0.377 \\
\bottomrule
\multicolumn{10}{l}{\textit{Notes:} All models include week dummies. Significance level: *** p$<$0.01, ** p$<$0.05, * p$<$0.10. Robust standard errors in parentheses.}
\end{tabular}}
\end{center}
\end{table}
%*-*-*-*-*-*-*-*-*-*-*-*-*-*-

To understand the magnitude of the estimated effect for different socioeconomic conditions, we use the estimated parameters and compute the difference between the lower and higher level of each indicator. For example, in the case of the socioeconomic strata, the values range between 1 and 6. The estimated effect before the lockdown for the lowest socioeconomic strata is $e^{(0.086*1)}=1.09$ and for the highest socioeconomic strata is $e^{(0.086*6)}= 1.68$. Thus, the estimated mobility flow of strata 1 with reference to strata 6 is $e^{(0.086)(1-6)}-1=-0.35$, which means that flows of the lowest socioeconomic strata 1 are 35\% smaller than those of the highest socioeconomic strata. After the lockdown, the estimated effect for the lowest socioeconomic strata is given by $e^{((0.086-0.173)*1)}=0.92$ and for the highest socioeconomic strata by $e^{((0.086-0.173)*6)}=0.59$. Therefore, the difference for the extreme values of socioeconomic strata is $e^{(0.086-0.173)(1-6)}-1=0.54$, which implies that the lowest socioeconomic strata reduce mobility a 54\% less than the population in the highest socioeconomic strata. Following the same procedure, we can estimate the effect of each independent variable. 

%-----------------------------
%\begin{figure}[h!]
%\begin{center}
%\includegraphics[width=\textwidth]{Figures.png}
%\caption{Changes}
%\label{fig:changes}
%\end{center}
%\end{figure}
%-----------------------------

\end{document}